\documentclass[prx,twocolumn,superscriptaddress,showpacs,floatfix,longbibliography]{revtex4-1}

\usepackage{amsfonts}
\usepackage{amsmath}
\usepackage{txfonts}
\usepackage{amssymb}
\usepackage{amsbsy} 
\usepackage{graphicx}
\usepackage{color}
\usepackage{mathdots} 
\usepackage{braket} 
\usepackage{lipsum}
\usepackage{hyperref}
\hypersetup{colorlinks=true, linkcolor=blue, anchorcolor=blue, citecolor=blue,urlcolor=blue}

\begin{document}
\title{Steady-state edge burst: From free-particle systems to interaction-induced phenomena }
\author{Yu-Min Hu}
\affiliation{ Institute for
	Advanced Study, Tsinghua University, Beijing,  100084, China }
\affiliation{Department of Physics, Princeton University, Princeton, NJ 08544, USA}
	
\author{Wen-Tan Xue}
\affiliation{ Institute for
	Advanced Study, Tsinghua University, Beijing,  100084, China }
\affiliation{ Department of Physics, National University of Singapore, Singapore 117542, Singapore }

\author{Fei Song}
\altaffiliation{songfeiphys@gmail.com}
\affiliation{ Institute for
	Advanced Study, Tsinghua University, Beijing, 100084, China }
 \affiliation{Kavli Institute for Theoretical Sciences, Chinese Academy of Sciences, Beijing, 100190, China }
 
\author{Zhong Wang}
 \altaffiliation{ wangzhongemail@tsinghua.edu.cn }
\affiliation{ Institute for
	Advanced Study, Tsinghua University, Beijing,  100084, China }

\begin{abstract}
	The interplay between the non-Hermitian skin effect and the imaginary gap of lossy lattices results in the edge burst, a boundary-induced dynamical phenomenon in which an exceptionally large portion of particle loss occurs at the edge. Here, we find that this intriguing non-Hermitian dynamical phenomenon can be exactly mapped into the steady-state density distribution of a corresponding open quantum system. Consequently, the bulk-edge scaling relation of loss probability in the edge burst maps to that of steady-state density. Furthermore, we introduce a many-body open-system model in which the two-body loss generates an interaction-induced non-Hermitian skin effect. Using the positive-$P$ method, we demonstrate the validity of the scaling relation for steady-state correlators. These results provide a unique perspective on the interaction-induced many-body non-Hermitian skin effect. Our predictions are testable in state-of-the-art experimental platforms.
\end{abstract}
\date{\today}
\maketitle
\section{Introduction}
Non-Hermitian systems exhibit various exotic phenomena that lack Hermitian counterparts  \cite{ Bergholtz2021RMP, Ashida2021}. A notable example is the non-Hermitian skin effect (NHSE), which describes the accumulation of bulk modes at the boundaries under open boundary conditions (OBC)  \cite{yao2018edge, yao2018chern, kunst2018biorthogonal, lee2018anatomy, alvarez2017,  Ghatak2019NHSE, xiao2020non, helbig2020generalized, Weidemann2020topological, wangwei2022non, Zhang2022ReivewOnNHSE, ding2022non, lin2023topological}. The NHSE profoundly reshapes the conventional topological band theory  \cite{yao2018edge,yao2018chern, Song2019real,Yokomizo2019,kunst2018biorthogonal, Yang2019Auxiliary,Deng2019, Kawabata2020nonBloch,liu2019second, Lee2020Unraveling,Yi2020,wang2022amoeba} and significantly alters the dynamical properties of non-Hermitian systems  \cite{Longhi2019nonBloch, Longhi2019Probing, Xiao2021nonBloch, Longhi2022self-healing, Longhi2022non, Wanjura2019,xue2021simple,hu2023greens,xue2022non}. Moreover, the Liouvillian NHSE in open quantum systems plays a crucial role in nonequilibrium steady states and relaxation dynamics  \cite{Song2019, Haga2021liouvillian, Liu2020helical, yang2022liouvillian, McDonald2022nonequilibrium,li2023manybody}.

Recently, a unique non-Hermitian dynamical phenomenon known as the \emph{edge burst} has been both theoretically proposed and experimentally observed  \cite{xue2022non,wang2021quantum,xiao2023observation}. This phenomenon highlights the surprisingly large loss probability at the boundary when a particle walks in a lossy lattice [see Fig.  \ref{fig:mapping}(c)]. The edge burst originates from a universal bulk-edge scaling relation, which is ascribed to the interplay between NHSE and gapless dissipative spectrum  \cite{xue2022non}. 

So far, the edge burst is seen as a transient phenomenon that occurs during a short period, and its detection requires a fine timing. It is therefore highly desirable to find its counterpart, if any, as steady states of open quantum systems, so that it persists for an arbitrarily long time. Meanwhile, the edge burst, as understood now, is a single-particle phenomenon.  In this work, we demonstrate that the dynamical phenomenon of the non-Hermitian edge burst can be accurately mapped to the properties of nonequilibrium steady states in bosonic open quantum systems (Fig. \ref{fig:mapping}). Concretely, the enhanced edge loss probability corresponds to a large steady-state particle density near the boundary of the open quantum system. Moreover, the universal bulk-edge scaling relation remains true for the steady-state density distribution. This offers a fresh correspondence between non-Hermitian dynamics and nonequilibrium steady states.

Unlike the single-particle edge burst, the steady-state edge burst here is a many-body phenomenon since the steady state contains many bosons. Furthermore, we have also included two-body loss, which amounts to having interaction effects in the quantum master equation  \cite{syassen2008strong,yan2013observation, Tomita2019dissipative, Gersema2021probing, Christianen2019photoinduced}. We found that the key features of the steady-state edge burst are robust, and also enriched, in such interaction effects. The on-site interactions introduce a new mechanism of many-body NHSE, which is different from widely studied non-Hermitian interacting models with asymmetric hoppings  \cite{Hamazaki2019non-hermitian_MBL, zhai2020Many-body, zhang2020skin, sen2020emergent, leeeunwoo2020many-body, lee2021many-body, kawabata2022many, alsallom2022fate, suthar2022non-Hermitian}. In this sense, the steady-state edge burst provides a promising avenue for detecting and exploring NHSE in dissipative many-body systems.

\section{Dynamical-steady correspondence}
\begin{figure*}
	\centering
\includegraphics[width=16cm]{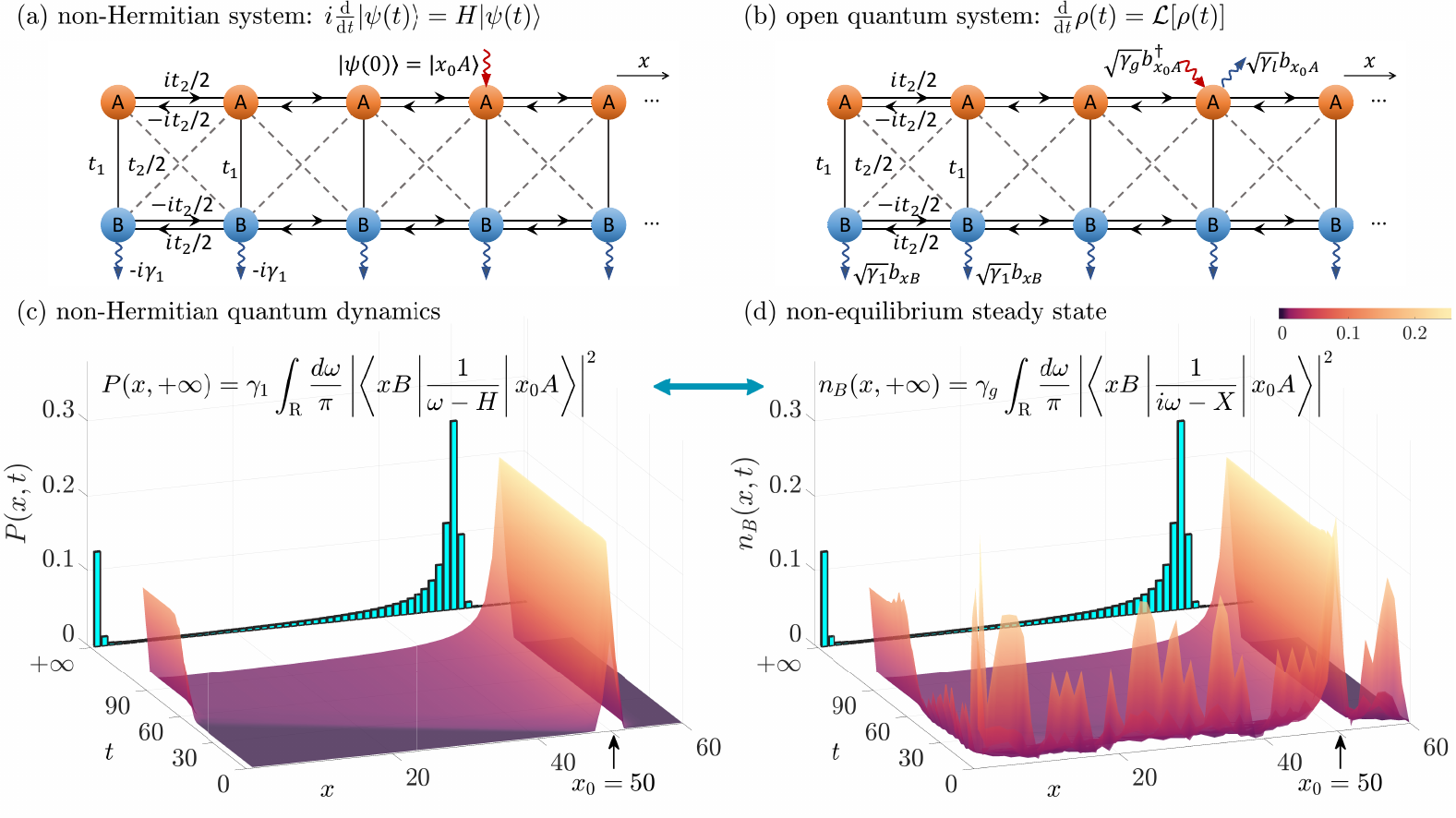}

	\caption{The dynamical-steady correspondence. (a) The Hamiltonian of the non-Hermitian dynamics. (b) The bosonic open quantum system. (c) The loss probability $P(x,t)$ generated by the initial state $\ket{x_0A}$. The blue histogram shows the total loss probability $P(x,+\infty)$. (d) The time evolution of $n_B(x,t)$ in the open quantum system with the initial state taken by randomly choosing each $n_A(x,0)$ from $\{0,1\}$. The blue histogram shows the steady-state density on $B $ sites $n_B(x,+\infty)$. Parameters: $t_1=0.8,t_2=1,\gamma_1=\gamma_g=\gamma_l=0.8,L=60,x_0=50$.}
	\label{fig:mapping}
\end{figure*}

We begin by elucidating a correspondence between non-Hermitian dynamics and steady states of open quantum systems. We consider a one-dimensional non-Hermitian system whose Bloch Hamiltonian is
\begin{eqnarray}
	H(k)=(t_1+t_2\cos k)\sigma_x+(t_2\sin k+i\frac{\gamma_1}{2})\sigma_z-i\frac{\gamma_1}{2}\sigma_0.
	\label{eq:nH_Hamiltonian}
\end{eqnarray}
In the above, $\sigma_{x,y,z}$ are Pauli matrices with $\sigma_z=+1\ (-1)$ denoting the $A$ ($B$) sublattice; $\sigma_0$ is the identity matrix. $t_1\ (t_2)$ is the intracell (intercell) hopping and $\gamma_1$ is the loss strength on $B$ sites [Fig.  \ref{fig:mapping}(a)]. This model is equivalent to the well-explored non-Hermitian Su-Schrieffer-Heeger model, with skin modes squeezed to the left edge  \cite{yao2018edge}.

We are interested in the non-Hermitian dynamics where a quantum walker described by $|\psi(t)\rangle$ evolves under $i\frac{\mathrm{d}}{\mathrm{d}t}\ket{\psi(t)}=H\ket{\psi(t)}$.  The real-space $H$ is generated by Eq. \eqref{eq:nH_Hamiltonian}.  In a length-$L$ chain under OBC, we use $x$ and $s\in \{\text{A, B}\}$ to represent the indices of unit cells and sublattices, respectively. Due to the inherent loss at sites $B$, the norm of the wavefunction decays as $\frac{\mathrm{d}}{\mathrm{d}t}\braket{\psi(t)|\psi(t)}=-2\gamma_1\sum_{x=1}^L|\braket{xB|\psi(t)}|^2$, which signifies the escape process of the walkers from sites $B$. Therefore, we define the site-resolved loss probability in the time interval $[0,t]$ as $P(x,t)=2\gamma_1\int_0^t\mathrm{d}t'|\braket{xB|\psi(t^\prime)}|^2$, and $P_x\equiv P(x,+\infty)$ represents the total loss probability at each $B $ site. All possibilities fulfill $\sum_{x=1}^L P_x=1$, which is ensured by the normalized condition of the initial state $\langle \psi(0)|\psi(0)\rangle =1$. We take the initial state to be $|\psi(0)\rangle=|x_0A\rangle$. Then, by utilizing the non-Hermitian Green's function  \cite{xue2021simple, hu2023greens, xue2022non}, we can express $P_x$ as
\begin{eqnarray}
	P_x\equiv P(x,+\infty)=\gamma_1\int_{-\infty}^{+\infty}\frac{\mathrm{d}\omega}{\pi}\left|\left\langle xB\left|\frac{1}{\omega-H} \right|x_0A\right\rangle\right|^2.
	\label{eq:integral_Px}
\end{eqnarray}

When $0<t_1\le t_2$, the Bloch spectrum of $H(k)$ in Eq.\eqref{eq:nH_Hamiltonian} closes its imaginary gap by touching the real axis, which means that the maximum imaginary part of the Bloch spectrum is zero \cite{xue2022non}. Under this condition, the numerical simulation [Fig.  \ref{fig:mapping}(c)] demonstrates that $P_x$ under OBC exhibits a remarkably pronounced peak at one edge, even though the starting point $x_0$ is far from this edge. This unexpected non-Hermitian dynamical phenomenon, called the \emph{edge burst}, was explained by the interplay between NHSE and imaginary gap closing  \cite{xue2022non}.  Recently, it has been observed on photonic platforms  \cite{xiao2023observation}.

The above non-Hermitian dynamical process is also expected in a bosonic open quantum system whose density matrix $\rho(t)$ follows the Lindblad master equation
\begin{eqnarray}
\frac{\mathrm{d}}{\mathrm{d}t}\rho=\mathcal{L}[\rho]=-i[\mathcal{H}_0,\rho]+\sum_\mu\left(2L_\mu\rho L_\mu^\dagger-\{L^\dagger_\mu L_\mu,\rho\}\right),\label{eq:single_lindblad}
\end{eqnarray}
where $\mathcal{L}$ is the Liouvillian superoperator. The coherent Hamiltonian is $\mathcal{H}_0=\sum_{ij}b_i^\dagger (H_0)_{ij}b_j$ with $H_0$ being the Hermitian part of $H$ in Eq. \eqref{eq:nH_Hamiltonian}, where $i$ and $j$ represent the indices $\{(x,s)\}$ and $b_i$ is the bosonic annihilation operator.  We consider single-particle loss $L_{1,xB}=\sqrt{\gamma_1}b_{xB}$ on each $B$ site. In this system, the quantum dynamics started from single-particle states are controlled by the effective Hamiltonian $\mathcal{H}_{\text{eff}}=\mathcal{H}_0-i\sum_{x=1}^L L_{1,xB}^\dagger L_{1,xB}=\sum_{ij}b_i^\dagger (H_{\text{eff}})_{ij} b_j$ where $H_{\text{eff}}$ coincides with the real-space Hamiltonian $H$ generated by Eq. \eqref{eq:nH_Hamiltonian}. Thus, the edge burst is anticipated from such dynamics.
 
Remarkably, we also unveil the edge burst in nonequilibrium steady states of open quantum systems. To make this clear, as shown in Fig.  \ref{fig:mapping}(b), we add new jump operators $L_{A}^{(l)}=\sqrt{\gamma_l}b_{x_0A}$ and $L_{A}^{(g)}=\sqrt{\gamma_g}b^\dagger_{x_0A}$ to the above system. Physically, this can be achieved by coupling the $A$ site at location $x_0$ with a reservoir that serves as a particle source. In this system, the time evolution of the correlator $\Delta_{ij}(t)=\langle b_i^\dagger b_j\rangle=\operatorname{Tr}[\rho(t) b_i^\dagger b_j]$ follows
\begin{eqnarray}
	\frac{\mathrm{d}}{\mathrm{d}t}\Delta(t)=X\Delta(t)+\Delta(t)X^\dagger+2M^{(g)},\label{eq:delta_evolution}
\end{eqnarray}
where $M^{(g)}_{ij}=\gamma_g\delta_{i,x_0A}\delta_{j,x_0A}$ and the damping matrix $X$ satisfies 
\begin{equation}
X_{ij}=(iH^*)_{ij}+(\gamma_g-\gamma_l)\delta_{i,x_0A}\delta_{j,x_0A}.\label{eq:single_damping}
\end{equation}
Here, $H$ is the real-space Hamiltonian of Eq.  \eqref{eq:nH_Hamiltonian}. The steady-state correlator $\Delta_{ss}=\Delta(t=+\infty)$ is given by $\frac{\mathrm{d}}{\mathrm{d}t}\Delta_{ss}= X\Delta_{ss}+\Delta_{ss}X^\dagger+2M^{(g)}=0$. Then, the steady-state density on $B$ sites has a formal expression (see detailed derivations in Ref.  \cite{McDonald2022nonequilibrium} and Appendix \ref{ap:steady})

 \begin{eqnarray} n_B(x,+\infty)\equiv\left(\Delta_{ss}\right)_{xB,xB}=\gamma_g\int_{-\infty}^{+\infty}\frac{\mathrm{d}\omega}{\pi}\left|\left\langle xB\left|\frac{1}{i\omega-X} \right|x_0A\right\rangle\right|^2.\label{eq:steady_density}
 \end{eqnarray} 
 
When taking balanced gain and loss $\gamma_g=\gamma_l$, the damping matrix $X$ in Eq.  \eqref{eq:single_damping} directly connects to the non-Hermitian Hamiltonian $H$ in real space as $X=iH^*$. Then, by comparing Eq.  \eqref{eq:integral_Px} and Eq.  \eqref{eq:steady_density}, we can easily find the equivalence between the loss probability $P_x$ and the steady-state density $n_B(x,+\infty)$ as $n_B(x,+\infty)=(\gamma_g/\gamma_1)P_x$.  Moreover, the normalization condition $\sum_{x=1}^LP_x=1$ is translated into a steady-state constraint $\sum_{x=1}^Ln_B(x,+\infty)={\gamma_g}/{\gamma_1}$, which is consistent with the steady-state condition $\frac{\mathrm{d}}{\mathrm{d}t}{N_{\text{tot}}}=0$ for the total particle number $N_{\text{tot}}=\sum_{x=1}^L\sum_{s\in\{A,B\}}\langle b_{xs}^\dagger b_{xs}\rangle$. This is a concrete example of dynamical-steady correspondence. With this correspondence, we can predict the edge burst in the steady state when $X$ has NHSE under OBC and a gapless dissipative spectrum under periodic boundary conditions (PBC). The latter condition means that the maximum real part of the PBC spectrum of $X$ is zero [see Fig. \ref{fig:single_scaling}(b)]. When the two conditions are both satisfied in the parameter regime $0<t_1\leq t_2$, starting from a random initial state, the system will eventually relax to the steady state with a prominent edge peak in the distribution of $n_B(x,+\infty)$ [Fig.  \ref{fig:mapping}(d)].
 

\section{ Bulk-edge scaling relation in non-equilibrium steady states}\label{sec:scaling_relation}
The dynamical phenomenon of the edge burst stems from the combination of NHSE and gapless dissipative spectrum. This combination gives rise to a universal bulk-edge scaling relation of the total loss probability  \cite{xue2022non}. Notably, this universal bulk-edge scaling relation also exists in the steady-state correlators of open quantum systems.

\begin{figure}
	\includegraphics[width=8.5cm]{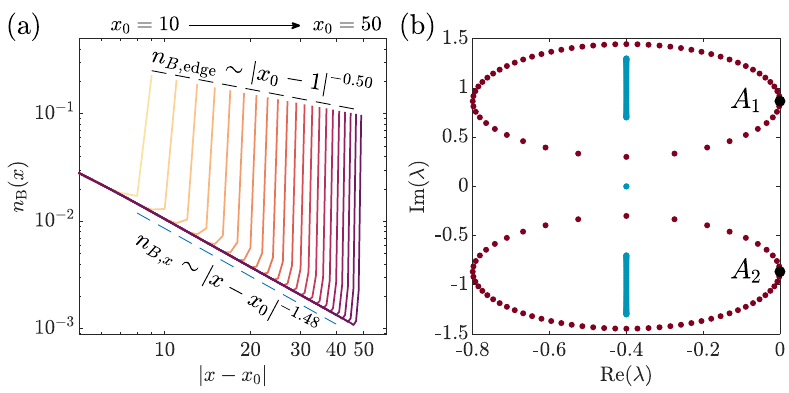}	
	\caption{The steady-state bulk-edge scaling relation in a quadratic open quantum system. (a) Steady-state values of $n_B(x)$ for different $x_0\in[10,50]$. The dashed lines represent numerical fittings of bulk (blue) and edge (black) values. (b) The Bloch spectrum (red) and OBC spectrum (blue) of $X$ with $\gamma_g=\gamma_l=0.8$. The black points in (b) are gapless eigenvalues of $X$ under PBC. Parameters: $t_1=0.5, \ t_2=1,\gamma_1=0.8,L=60$. }\label{fig:single_scaling}
\end{figure}

As shown in Fig.  \ref{fig:single_scaling}, when the dissipative spectrum is gapless, the steady-state density $n_B(x)$ in the bulk region follows an algebraic (power-law) decay as $n_{B}(x)\sim|x-x_0|^{-\alpha_b}$ for $x<x_0$. Similarly, at the left edge,  $n_{B,\text{edge}}\equiv n_{B}(x=1)$ also exhibits an algebraic decay as $n_{B,\text{edge}}\sim|x_0-1|^{-\alpha_e}$, where $|x_0-1|$ is the distance from the edge to the pumping site.  The edge burst shown in Fig.  \ref{fig:mapping}(d) originates from a bulk-edge scaling relation between the two scaling exponents:
\begin{equation}
	\alpha_e=\alpha_b-1.
	\label{eq:scaling_relation}
\end{equation}

The presence of NHSE and gapless dissipative spectrum in our open quantum system is crucial for the bulk-edge scaling relation in steady states. On the one hand, NHSE appears in an intuitive way. The phase difference between hoppings in $A$ and $B$ chains generates motions in opposite directions, and loss processes suppress the rightward motion on the $B$ chain, causing a net leftward motion. On the other hand, when $0<t_1\le t_2$ and $\gamma_l=\gamma_g=0$, the gapless dissipative spectrum under PBC explicitly comes from the dark states $\ket{\phi_n}=({1}/{\sqrt{n!}})(b_{k_0}^\dagger)^n\ket{0}$ with $b_{k_0}^\dagger =({1}/{\sqrt{L}})\sum_{x=1}^Le^{ik_0x}b^\dagger_{xA}$, where the momentum $k_0$ satisfies the decoupling condition $t_1+t_2\cos k_0=0$. With zero amplitude on sites $B$, these dark states fulfill $L_{1,xB}\ket{\phi_n}=0$. They form a dark space $\rho_{mn}=\ket{\phi_m}\bra{\phi_n}$ with purely imaginary Liouvillian spectrums $\mathcal{L}[\rho_{mn}]=-i(m-n)t_2\sin k_0\rho_{mn}$.  As local impurities, $L_A^{(g)}$ and $L_A^{(l)}$ at $x_0$ perturb the eigenoperators $\rho_{mn}$ locally, contributing negligibly to the PBC spectrum when $L\to+\infty$. As long as the localized impurity state is stable, the Liouvillian PBC spectrum remains gapless.  The points $A_{1,2}$ in Fig.  \ref{fig:single_scaling}(b) represent the gapless eigenvalues of the PBC damping matrix $X$ with $\gamma_g=\gamma_l\ne0$, which are related to the dark states of the Liouvillian. 

The bulk-edge scaling relation has a quantitative explanation  \cite{xue2022non}. On an infinite chain, the interplay between NHSE and gapless dissipative spectrum induces a slow algebraic decay $n_B^\infty(x,+\infty)\sim|x-x_0|^{-\alpha_b}$ for $x<x_0$. However,  a finite system introduces boundaries at $x=1$, and the large residual portion of particles $n_B^{\infty}(x,+\infty)$ for $x\le 0$ needs to be redistributed into the finite system. Due to NHSE-induced localization, these residual particles will accumulate at the left edge $x=1$.  With constraints $\sum_{x=-\infty}^{+\infty} n_B^\infty (x, +\infty)=\sum_{x=1}^{L} n_B (x, +\infty)=\gamma_g/\gamma_1$, the edge density $n_{B,\text{edge}}=n_B(x=1,+\infty)$ can be estimated as $n_{B,\text{edge}}\sim\sum_{x\le1}|x-x_0|^{-\alpha_b}\sim\int_{-\infty}^1\mathrm{d}x|x-x_0|^{-\alpha_b}\sim|x_0-1|^{-\alpha_b+1}$. Consequently, the ratio between $n_{B,\text{edge}}$ and the nearby bulk density is proportional to $|x_0-1|$, leading to a pronounced steady-state edge burst. Therefore, we establish the steady-state bulk-edge scaling relation as Eq. \eqref{eq:scaling_relation}. The numerical results in Fig.  \ref{fig:single_scaling}(a) closely match the theoretical values $\alpha_b=1.5$ and $\alpha_e=0.5$  \cite{xue2022non}.  The detailed calculation is presented in Appendix  \ref{ap:scaling}.

Although the aforementioned dynamical-steady correspondence is exact in bosonic systems when $\gamma_g=\gamma_l$, we emphasize that the steady-state edge burst still exists when $\gamma_g\ne\gamma_l$. Compared with the balanced situation, the imbalanced gain and loss at one $A$ site introduce a local impurity in the damping matrix [see Eq. \eqref{eq:single_damping}]. This additional impurity changes the overall factors in the distribution of $n_B(x,+\infty)$, but it will never modify the algebraic behaviors of $n_B(x,+\infty)$ as long as the imbalance $\delta \gamma=\gamma_g-\gamma_l$ is below a threshold to make the steady state stable. The detailed analysis is shown in Appendix  \ref{ap:imbalance}.  This scenario can also be extended to predict the steady-state edge burst in fermionic open quantum systems. With the same setups as in Fig.  \ref{fig:mapping}(b), the single-particle fermionic gain and loss dissipators at one $A$ site contribute a local impurity term $-(\gamma_g+\gamma_l)\delta_{i,x_0A}\delta_{j,x_0A}$ to the damping matrix in Eq.  \eqref{eq:single_damping}. Due to fermionic statistics, this term always induces a stable steady state. According to the above discussion, this local impurity term does not affect the presence of the steady-state edge burst in fermionic systems.

\section{ Steady-state edge burst induced by two-body loss}
The bulk-edge scaling relation of steady states is remarkably general in open quantum systems, as long as both NHSE and gapless dissipative spectrum are present.  So far, our discussion focuses on non-interacting systems. A natural question arises: Does the edge burst still persist in the presence of many-body effects? We find that the steady-state edge burst, along with the accompanying universal bulk-edge scaling relation, is not only robust, but even enriched in dissipative many-body systems. To illustrate this point, we introduce a many-body open quantum system with experimentally accessible two-body loss $L_{2,xB}=\sqrt{\gamma_2}b_{xB}b_{xB}$  \cite{syassen2008strong,yan2013observation, Tomita2019dissipative, Gersema2021probing, Christianen2019photoinduced}, which replaces $L_{1,xB}=\sqrt{\gamma_1}b_{xB}$ in Fig.  \ref{fig:mapping}(c). 

The two-body loss plays a twofold role. First, it can bring many-body NHSE. The two-body loss dampens the unidirectional motion along the $ B$ chain, which results in many-body NHSE toward a preferable direction. This mechanism of interaction-induced NHSE differs from many-body NHSE induced by asymmetric hoppings  \cite{Hamazaki2019non-hermitian_MBL, zhai2020Many-body, zhang2020skin, sen2020emergent, leeeunwoo2020many-body, lee2021many-body, kawabata2022many, alsallom2022fate, suthar2022non-Hermitian}. Second,  the dark states $\ket{\phi_n}=({1}/{\sqrt{n!}})(b_{k_0}^\dagger)^n\ket{0}$ are immune to the presence of two-body loss since $L_{2,xB}\ket{\phi_n}=0$. Although determining the full many-body spectrum is challenging, we deduce that these dark states still induce a gapless dissipative spectrum, whose maximal real part approaches zero as the system size grows. These dark states also imply that the particles on the $A$ chain can travel considerable distances without any dissipation before either jumping into $B$ sites or being scattered by the boundary.  

The interplay between many-body NHSE and gapless dissipative spectrum indicates the steady-state edge burst with interactions. To illustrate this point, we examine the steady-state correlators by a phase-space method called the positive-$P$ method  \cite{carmichael1999statistical,gardiner2004quantum, Drummond_1980,smith1989simulation, Gilchrist1997positive, Deuar2006First,deuar2021fully}. This approach maps the Lindblad master equation into a Fokker-Planck equation of a probability distribution function in the phase space of bosons. Equivalently, the dissipative many-body dynamics is obtained by simulating a set of stochastic differential equations of the phase-space variables and averaging over random trajectories.  By performing long-time simulations, we can obtain steady-state correlators. Numerical details are presented in Appendix  \ref{ap:positive-P}. 

We consider balanced gain and loss $\gamma_g=\gamma_l$ hereafter. Different from the quadratic case,  the steady-state condition $\frac{\mathrm{d}}{\mathrm{d}t}N_{\text{tot}}=0$ brings a new constraint on the four-point correlators $\sum_{x=1}^L\langle b_{xB}^\dagger b_{xB}^\dagger b_{xB} b_{xB}\rangle=\gamma_g/(2\gamma_2)$. As a result, we examine the steady-state distributions of the $B$-site particle number $n_B(x)=\operatorname{Tr}(\rho_{ss} b_{xB}^\dagger b_{xB})$ as well as the four-point correlator: 
\begin{equation}
	C_B(x)=\operatorname{Tr}(\rho_{ss} b_{xB}^\dagger b_{xB}^\dagger b_{xB} b_{xB}),
\end{equation}
where $\rho_{ss}$ is the steady-state density matrix.

\begin{figure}
\centering
\includegraphics[width=8.5cm]{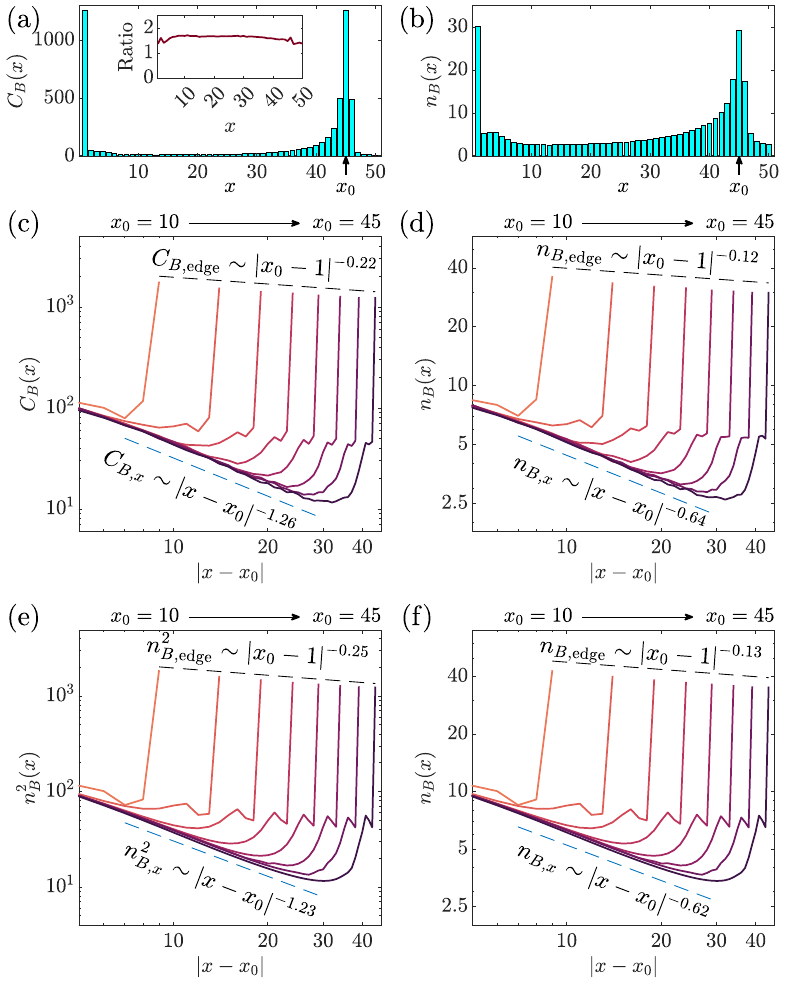}
	\caption{The steady-state bulk-edge scaling relation with two-body loss. (a) $C_B(x)$ and (b) $n_B(x)$ for $x_0=45$.  The inset in (a) shows the ratio $C_B(x)/n_B^2(x)$.  (c) $C_B(x)$ and (d) $n_B(x)$ for $x_0\in\{10,15,\cdots,45\}$. Results in (a)--(d) are obtained by the positive-$P$ method. We take the average over $10^4$ samples of stochastic trajectories. (e) $n_B^2(x)$ and (f) $n_B(x)$ obtained from the mean-field approximation. Dashed lines in (c)--(f) are given by numerical fittings of bulk (blue) and edge (black) values. Parameters: $t_1=0.8,t_2=1,\gamma_1=0,\gamma_2=0.01,\gamma_l=\gamma_g=100,L=50$.}\label{fig:P_rep_scaling}
\end{figure}

Figures  \ref{fig:P_rep_scaling}(a)--3(d) illustrate the steady-state correlators obtained by the positive-$P$ method, where we set $0<t_1<t_2$. First, the asymmetric distributions around $x_0$ indicate the presence of many-body NHSE [Figs.  \ref{fig:P_rep_scaling}(a), (b)]. Furthermore, we observe the steady-state edge burst with prominent peaks on the left edge [Figs.  \ref{fig:P_rep_scaling}(a), (b)]. Quantitatively, the bulk values of $C_B(x)$ and $n_B(x)$ exhibit algebraic scaling behaviors with respect to $|x-x_0|$ where $x<x_0$, and the edge values also decay algebraically with respect to $|x_0-1|$ [Figs.  \ref{fig:P_rep_scaling}(c), (d)]. 

These numerical results elucidate that the bulk and edge scaling factors of $C_B(x)$ and $n_B(x)$ follow two different relations, which are $\alpha_b-\alpha_e\approx 1.04$ for $C_B(x)$ and $\alpha_b-\alpha_e\approx 0.52$ for $n_B(x)$. These scaling relations can be  explained as follows. With the steady-state constraint $\sum_{x=1}^LC_B(x)={\gamma_g}/(2\gamma_2)$,   the analysis presented in Sec. \ref{sec:scaling_relation}, which leads to Eq.\eqref{eq:scaling_relation}, can likewise be extended to the case of two-body loss. As a result, the algebraic decay of $C_B(x)$ and the NHSE-induced boundary accumulation directly lead to the bulk-edge scaling relation $\alpha_b-\alpha_e=1$. Additionally, compared with $C_B(x)$, the half-scaling behavior of $n_B(x)$ is associated with an important numerical observation $C_B(x)/n_B^2(x)\approx O(1)$ [see the inset in Fig. \ref{fig:P_rep_scaling}(a)].

Inspired by this observation, we come up with mean-field explanations for the numerical findings (see details in Appendix  \ref{ap:meanfield}). To mimic the effect of two-body loss $L_{2,xB}=\sqrt{\gamma_2}b_{xB}b_{xB}$, we employ the mean-field linear bosonic loss operators $L_{\text{MF},xB}=\sqrt{2\gamma_2n_{B,\text{MF}}(x,t)}b_{xB}$, where the loss strength is determined by local particle density $n_{B,\text{MF}}(x,t)=\operatorname{Tr}(\rho(t) b_{xB}^\dagger b_{xB})=\Delta_{xB,xB}(t)$. These empirical mean-field linear loss operators introduce nonlinearity to Eq. \eqref{eq:delta_evolution} by replacing the elements $X_{xB,xB}=-\gamma_1$ with $(X_{\text{MF}})_{xB,xB}=-2\gamma_2\Delta_{xB,xB}(t)$. $X_{\text{MF}}$ denotes the mean-field damping matrix. During the non-linear evolution, $X_{\text{MF}}$ under PBC also possesses gapless eigenvalues with the corresponding states having zero amplitude at the $B$ sites [similarly to $A_{1,2}$ in Fig. \ref{fig:single_scaling}(b)]. In this context, particles can propagate long distances on the $A$ chain before escaping from the $B$ sites or reaching the left edge. Similarly to the noninteracting cases, this process induces a slowly decaying algebraic distribution of the steady-state bulk density.  Quantitatively, at the mean-field level, the steady-state density in the bulk region scales as $n_{B,\text{MF}}(x)\sim\tilde\gamma(x)^{-1.5}|x-x_0|^{-1.5}$ with $\tilde\gamma(x)\sim 2\gamma_2n_{B,\text{MF}}(x)$  contributed by the density-dependent loss in $X_{\text{MF}}$. Consequently, the self-consistent relation $n_{B,\text{MF}}(x)\sim n_{B,\text{MF}}(x)^{-1.5}|x-x_0|^{-1.5} $ results in the scaling $n_{B,\text{MF}}(x)\sim|x-x_0|^{-0.6}$ . With the mean-field constraint $\sum_{x=1}^L n_{B,MF}^2(x)={\gamma_g}/(2\gamma_2)$, a similar explanation below Eq. \eqref{eq:scaling_relation} leads to $n^2_{B,\text{edge}}(x_0)\sim|x_0-1|^{-0.2}$ and $n_{B,\text{edge}}(x_0)\sim|x_0-1|^{-0.1}$, which is in good agreement with the numerical results obtained from the positive-$P$ method [Fig.  \ref{fig:P_rep_scaling}].

\section{ Discussion} 

In this work, we uncovered a correspondence between edge burst phenomena in non-Hermitian quantum dynamics and steady-state correlators in open quantum systems. This correspondence provides experimental feasibility to explore dynamical phenomena by preparing suitable steady states. Moreover, in dissipative many-body systems, we discovered a new mechanism of many-body NHSE, which enriches our understanding of the universal bulk-edge scaling relation of the steady-state edge burst in dissipative interacting systems. Importantly, our findings demonstrated that steady-state correlators contain valuable information on many-body NHSE, offering an experimentally accessible route to detect NHSE in dissipative many-body systems. Exploring other types of dualities between non-Hermitian dynamics and nonequilibrium steady states in many-body open quantum systems would be an interesting topic for future research.

\section*{Acknowledgment}

We thank He-Ran Wang and Zhou-Quan Wan for helpful discussions. This work is supported by NSFC under Grant No. 12125405.

\appendix

\section{Steady states in quadratic open quantum systems}\label{ap:steady}
The density matrix of a quadratic open quantum system follows the Lindblad master equation as Eq.  \eqref{eq:single_lindblad} with a quadratic Hamiltonian $\mathcal{H}_0=\sum_{ij}b_i^\dagger (H_0)_{ij}b_j$ and linear Lindblad operators $L_\mu^{(g)}=\sum_iD^{(g)}_{\mu,i}b_i^\dagger$ and $L_\mu^{(l)}=\sum_jD^{(l)}_{\mu,j}b_j$. The operators $b_i,b_i^\dagger$ can be either bosonic (i.e., $[b_i,b_j]=[b_i^\dagger,b_j^\dagger]=0$ and $[b_i,b_j^\dagger]=\delta_{ij}$) or fermionic (i.e., $\{b_i,b_j\}=\{b_i^\dagger,b_j^\dagger\}=0$ and $\{b_i,b_j^\dagger\}=\delta_{ij}$). A special property of quadratic open quantum systems is that their steady states $\rho_{ss}\equiv\rho(t \rightarrow +\infty)$ are always Gaussian. In other words, the information contained in the steady state can be completely extracted from its two-point correlator $(\Delta_{ss})_{ij}=\operatorname{Tr}[\rho_{ss} b_i^\dagger b_j]$. As we have mentioned in the main text, the steady-state correlator $\Delta_{ss}$ satisfies a matrix equation
\begin{equation}
  X  \Delta_{ss} +\Delta_{ss}X^\dagger+2M^{(g)}=0,\label{seq:Sylvester}
\end{equation}
where the so-called damping matrix is given by $X=i H_0^T+M^{(g)}-(M^{(l)})^T$ in a bosonic system (or $X=i H_0^T-M^{(g)}-(M^{(l)})^T$ in a fermionic system), with $M^{(g)}=\sum_\mu (D^{(g)}_\mu)^\dagger D^{(g)}_\mu$ and $ M^{(l)}=\sum_\mu (D^{(l)}_\mu)^\dagger D^{(l)}_\mu$. This equation can be solved as follows. First, we can transform the matrices $\Delta_{ss}$ and $M^{(g)}$ into two supervectors as $|\Delta_{ss})=\sum_{ij}(\Delta_{ss})_{ij}\ket{i}\otimes\ket{j}$ and $|M^{(g)})=\sum_{ij}M^{(g)}_{ij}\ket{i}\otimes\ket{j}$. Through this transformation, the matrix equation Eq.  \eqref{seq:Sylvester} is converted to
\begin{equation}
(X\otimes I+I\otimes X^*)|\Delta_{ss})=-2|M^{(g)}). \label{seq:superstate}
\end{equation}
Then, $|\Delta_{ss})$ can be obtained via calculating the inverse of $X\otimes I+I\otimes X^*$. After diagonlizing the damping matrix as $X=\sum_n\lambda_n\ket{nR}\bra{nL}$, the inverse of $X\otimes I+I\otimes X^*$ equals to
\begin{equation}
\sum_{m,n}\ket{mR}\otimes\ket{nR^*}\frac{1}{\lambda_m+\lambda^*_n}\bra{mL}\otimes\bra{nL^*}.
\end{equation}
Inserting this back to Eq.   \eqref{seq:superstate}, we achieve a formal expression of $|\Delta_{ss})$. This expression can be further simplified by considering the equality 
\begin{equation}
	\frac{1}{\lambda_m+\lambda^*_n}=\int_{-\infty}^{+\infty}\frac{\mathrm{d}\omega}{2\pi}\frac{1}{i\omega-\lambda_m}\times\frac{1}{i\omega+\lambda_n^*}. \label{seq:lambda_equality}
\end{equation}
This equality holds when the damping matrix only has eigenvalues with nonpositive real parts, namely $\text {Re}(\lambda_n)\leq 0$. This condition is also necessary
for the existence of a well-defined steady state. With the equality Eq.  \eqref{seq:lambda_equality} and mapping $|\Delta_{ss})$ back to its matrix representation, we derive that
\begin{equation}
	\begin{split}
		(\Delta_{ss})_{ij}&=-2\sum_{k,l,m,n}M^{(g)}_{kl}\int_{-\infty}^{+\infty}\frac{\mathrm{d}\omega}{2\pi}\frac{\braket{i|mR}\braket{mL|k}\braket{j|nR^*}\braket{nL^*|l}}{(i\omega-\lambda_m)(-i\omega+\lambda_n)^*}\\
		&=2\sum_{k,l}M^{(g)}_{kl}\int_{-\infty}^{+\infty}\frac{\mathrm{d}\omega}{2\pi}\left\langle i\left|\frac{1}{i\omega-X}\right|k\right\rangle\left\langle j\left|\left(\frac{1}{i\omega-X}\right)^*\right|l\right\rangle \\
		&=\int_{-\infty}^{+\infty}\frac{\mathrm{d}\omega}{\pi}\left\langle i\left|\frac{1}{i\omega-X}M^{(g)}\left(\frac{1}{i\omega-X}\right)^\dagger\right|j\right\rangle. 
	\end{split}\label{seq:slyvester_solution}
\end{equation}
This result leads to Eq.  \eqref{eq:steady_density} in the main text. To see more applications of Eq.  \eqref{seq:slyvester_solution}, one can consult Ref.  \cite{McDonald2022nonequilibrium}.
\section{Algebraic scalings in steady states}\label{ap:scaling}
\begin{figure}
	\centering
	\includegraphics[width=8.5cm]{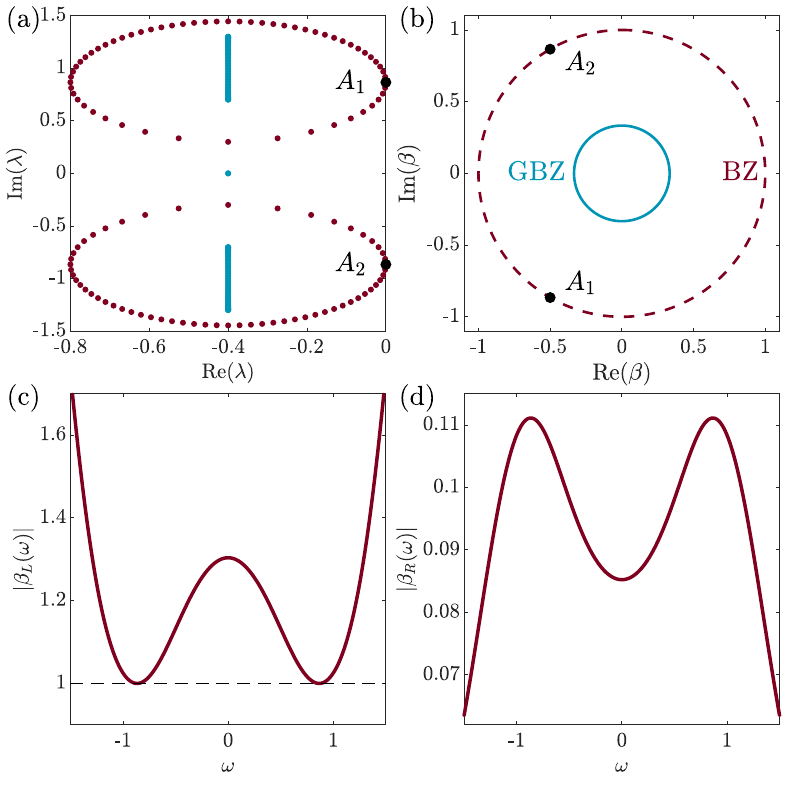}
	\caption{(a) The PBC spectrum (red) and OBC spectrum (blue) of the damping matrix $X$ in Eq.  \eqref{seq:balanced_damping_matrix}. We set $L=60$. $A_1$ and $A_2$ are the gapless points $\lambda=i\omega_0$ on the imaginary axis. (b) Brillouin zone (red dashed line) and generalized Brillouin zone (blue solid line). $A_1$ and $A_2$ are $\beta_L(\omega_0)$ with $\omega_0=\pm\sqrt{t_2^2-t_1^2}$ respectively. (c) $|\beta_L(\omega)|$ changes as $\omega$. (d) $|\beta_R(\omega)|$ changes as $\omega$. Parameters: $t_1=0.5,\ t_2=1,\ \gamma_1=\gamma_g=\gamma_l=0.8$.}
	\label{sfig:pbc_gapless}
\end{figure}

In the main text, we consider a quadratic bosonic open quantum system with Hamiltonian
\begin{equation}
\begin{split}
H_0=&\sum_{x=1}^L t_1\left(b_{xA}^\dagger b_{xB}+h.c.\right)\\
	&+\sum_{x=1}^{L-1}\frac{t_2}{2}\left(b_{xA}^\dagger b_{(x+1)B}+b_{(x+1)A}^\dagger b_{xB}+h.c.\right)\\
 &+\sum_{x=1}^{L-1}\frac{it_2}{2}\left(b_{(x+1)A}^\dagger b_{xA}-b_{(x+1)B}^\dagger b_{xB}-h.c.\right).
 \end{split}
\end{equation}
and three types of jump operators
\begin{equation}
	\begin{split}
		L_{1,xB}&=\sqrt{\gamma_1}b_{xB},\quad x=1,2,\cdots,L;\\
		L_{A}^{(l)}&=\sqrt{\gamma_l}b_{x_0A};\\
		 L_{A}^{(g)}&=\sqrt{\gamma_g}b_{x_0A}^\dagger.
	\end{split}
\end{equation}
When $\gamma_l=\gamma_g$, the damping matrix of this model has a momentum-space expression
\begin{equation}
	X(k)=i(t_1+t_2\cos k)\sigma_x-i(t_2\sin k+i\frac{\gamma_1}{2})\sigma_z-\frac{\gamma_1}{2}\sigma_0.\label{seq:balanced_damping_matrix}
\end{equation}
This damping matrix has diverse spectrums under PBC and OBC [Fig.~ \ref{sfig:pbc_gapless}(a)], which is a fingerprint of NHSE. Furthermore, the dissipative gap closes under PBC when $t_1<t_2$. The dissipative gap closing can induce algebraic (power-law) scalings in steady states. For example, the steady-state density distribution $n_B(x)$, which is determined by Eq. \eqref{eq:steady_density}, follows the algebraic decay in the bulk region of an OBC chain. Subsequently, we will give detailed derivations of it.

When $x$ is far from the edge, the matrix element $\langle xB\left|(i\omega-X)^{-1}\right|x_0A\rangle$ in Eq.  \eqref{eq:steady_density} can be expressed as
\begin{equation}
	\left\langle xB\left|\frac{1}{i\omega-X}\right|x_0A\right\rangle=\oint_{C} \frac{\mathrm{d} \beta}{2 \pi i \beta} \beta^{x-x_{0}}\left(\frac{1}{i\omega-X(\beta)}\right)_{B A}.\label{seq:contour_integral}
\end{equation}
where $X(\beta)\equiv X(k)|_{e^{ik}\to\beta}$ and the contour $C$ takes the generalized Brillouin zone (GBZ)  \cite{xue2021simple}. Here, the GBZ is a circle on the complex plane with a radius $|\beta|=\sqrt{|(t_1-\gamma_1/2)/(t_1+\gamma_1/2)|}$ [Fig.  \ref{sfig:pbc_gapless}(b)]. The damping matrix in Eq.  \eqref{seq:balanced_damping_matrix} provides
\begin{equation}
	\left(\frac{1}{i\omega-X(\beta)}\right)_{B A}=\frac{i(t_1+\frac{t_2}{2}(\beta+\beta^{-1}))}{\det[i\omega -X(\beta)]}.
\end{equation}
Consequently, the poles of the integral in Eq.~( \ref{seq:contour_integral}) are contributed by the two roots of $\det[i\omega -X(\beta)]=0$, which we denote as $\beta_L(\omega)$ and $\beta_R(\omega)$. They obey $|\beta_L(\omega)|>|\beta_R(\omega)|$ for $\omega\in\mathbb{R}$ when $t_1<t_2$. Due to the fact that $|\beta_L(\omega)\beta_R(\omega)|=|(t_1-\gamma_1/2)/(t_1+\gamma_1/2)|$,  $\beta_L(\omega)$ and $\beta_R(\omega)$ reside outside and inside GBZ, respectively. Thus,  Eq.~ \eqref{seq:contour_integral} is determined by the residue at $\beta_R(\omega)$ ($\beta_L(\omega)$) when $x>x_0$ ($x<x_0$), leading to 
\begin{eqnarray}
	n_B(x)=\begin{cases}
		\gamma_g\int_{-\infty}^{+\infty}\frac{\mathrm{d}\omega}{\pi}|f_R(\omega)|^2|\beta_R(\omega)|^{2(x-x_0)},\quad x>x_0\\
		\gamma_g\int_{-\infty}^{+\infty}\frac{\mathrm{d}\omega}{\pi}|f_L(\omega)|^2|\beta_L(\omega)|^{2(x-x_0)},\quad x<x_0
	\end{cases}\label{seq:bulk_B_integral}
\end{eqnarray}
The coefficients $f_{L}(\omega)$ and $f_{R}(\omega)$ are given by
\begin{equation}
f_{L / R}(\omega)=\lim _{\beta \to\beta_{L / R}}\left(\beta-\beta_{L / R}(\omega)\right) \frac{i(t_1+\frac{t_2}{2}(\beta+\beta^{-1}))}{\beta\det[i\omega+0^+ -X(\beta)]}.
\end{equation}
The integrals in Eq.  \eqref{seq:bulk_B_integral} are dominated by the neighbor of the frequency $\omega_0$ that gives $\max_{\omega\in\mathbb{R}} |\beta_{R}(\omega)|$ or $\min_{\omega\in\mathbb{R}}|\beta_{L}(\omega)|$. These integrals can be evaluated by expanding $f_{L/R}(\omega)$ and $\beta_{L/R}(\omega)$ to the leading order of $\delta \omega=\omega-\omega_0$.  In the regime $0<t_1<t_2$,  the dissipative gap closing accompanies $\min_{\omega\in\mathbb{R}}|\beta_{L}(\omega)|=|\beta_{L}(\omega_0)|=1$ at $\omega_0=\pm \sqrt{t_2^2-t_1^2}$ [Fig.  \ref{sfig:pbc_gapless}(c)]. For later convenience, we focus on $\omega_0=\sqrt{t_2^2-t_1^2}$ and present the expressions $\beta_{L}\left(\omega_{0}\right)=-\frac{t_{1}}{t_{2}}-i \frac{\sqrt{t_{2}^{2}-t_{1}^{2}}}{t_{2}}$ and $ \beta_{R}\left(\omega_{0}\right)=\frac{t_{1}-\gamma_1 / 2}{t_{1}+\gamma_1 / 2}\left(-\frac{t_{1}}{t_{2}}+i \frac{\sqrt{t_{2}^{2}-t_{1}^{2}}}{t_{2}}\right)$. The leading-order expansions around $\omega_0$ are  \cite{xue2022non}
\begin{equation}
	\begin{split}
f_L(\omega_0+\delta\omega)&\approx-\frac{\sqrt{t_2^2-t_1^2}}{t_1^2\left(2\sqrt{t_2^2-t_1^2}-i\gamma_1\right)}\delta\omega,\\
\ln|\beta_L(\omega_0+\delta\omega)|&\approx{\frac{\gamma_1  (t_{2}^{2}-t_{1}^{2})}{t_{1}^{3}\left( 4t_{2}^{2}-4t_{1}^{2}+\gamma_1^2\right)}(\delta\omega)^2}.
\end{split}\label{seq:delta_omega_expansion}
\end{equation}
In the second line, we define the coefficient before $(\delta\omega)^2$ as $K=\frac{\gamma_1  (t_{2}^{2}-t_{1}^{2})}{t_{1}^{3}\left( 4t_{2}^{2}-4t_{1}^{2}+\gamma_1^2\right)}$. Finally, the steady-state density distribution $n_B^\infty(x)$ for $x<x_0$ on an infinite chain is given by
\begin{equation}
	\begin{split}
		n_B^\infty(x)\sim\gamma_g\int\mathrm{d}(\delta\omega)(\delta\omega)^2e^{-2K(\delta\omega)^2|x-x_0|}\sim\gamma_g|x-x_0|^{-\frac{3}{2}}.
	\end{split}\label{seq:delta_omega_integral}
\end{equation}

This expression indicates the algebraic scaling of $n_B(x)$  and tells us the bulk scaling exponent $\alpha_b=1.5$. At the same time, as a result of the bulk-edge scaling relation $\alpha_e=\alpha_b-1$,  the edge scaling exponent $\alpha_e=0.5$. These theoretical predictions are verified by numerical simulations. 

In addition,  $|\beta_R(\omega)|$ also reaches the maximal value $\max_{\omega\in\mathbb{R}} |\beta_{R}(\omega)|=|\beta_R(\omega_0)|<1$ at  $\omega_0=\pm \sqrt{t_2^2-t_1^2}$ [Fig.  \ref{sfig:pbc_gapless}(d)]. Therefore, when $x>x_0$, the steady-state density decays exponentially $n_B(x)\sim |\beta_R(\omega_0)|^{2(x-x_0)}$.

\section{Imbalanced gain and loss}\label{ap:imbalance}

As discussed in the main text, the balanced incoherent pumping ($\gamma_g=\gamma_l$) at one $A$ site is crucial for the exact dynamical-steady correspondence. Due to this correspondence, the bulk-edge scaling relation established in the non-Hermitian edge burst can be revealed in the steady-state edge burst of a corresponding bosonic open quantum system. In this section, we show the robustness of steady-state edge burst in the presence of imbalanced gain and loss ($\gamma_g\ne\gamma_l)$. 

The imbalanced gain and loss introduce a local impurity to the damping matrix in Eq.  \eqref{seq:balanced_damping_matrix}. The modified damping matrix $\tilde X$ in real space is given by
\begin{equation}
	\tilde X=X+\delta\gamma\ket{x_0A}\bra{x_0A}.\label{seq:imbalanced_damping_matrix}
\end{equation}
$\delta\gamma=\gamma_g-\gamma_l$ represents the imbalance between gain and loss. $X$ is the real-space version of the balanced damping matrix defined in Eq.  \eqref{seq:balanced_damping_matrix}.

Due to the imbalance, the evolution of the total number of particles is given by $
	\frac{\mathrm{d}N_{\text{tot}}}{\mathrm{d}t}=-2\gamma_1\sum_x\langle b_{xB}^\dagger b_{xB}\rangle+2\delta\gamma\langle b_{x_0A}^\dagger b_{x_0A}\rangle+2\gamma_g$. As a result, the constraint on the steady-state density becomes
 \begin{equation}
 	\gamma_1\sum_x\langle b_{xB}^\dagger b_{xB}\rangle=\delta\gamma\langle b_{x_0A}^\dagger b_{x_0A}\rangle+\gamma_g.
 \end{equation}
 This equation indicates that the total number of particles at the $B$ sites $\sum_x\langle b_{xB}^\dagger b_{xB}\rangle$ is still a finite number as long as $\langle b_{x_0A}^\dagger b_{x_0A}\rangle$ is finite. The finite $\langle b_{x_0A}^\dagger b_{x_0A}\rangle$ is guaranteed by the stability of the imbalance-induced impurity state, which requires that $\delta\gamma$ is below a threshold $\delta \gamma_c$. Under this condition, an algebraic scaling of $n_B(x)\equiv\langle b_{xB}^\dagger b_{xB}\rangle$ can still induce the steady-state edge burst. 

To show the robustness of the steady-state edge burst against the local imbalance, we calculate the modified Green's function:
\begin{equation} 
	\tilde G(\omega)=\frac{1}{i\omega-\tilde X}=\frac{1}{i\omega-X-\delta\gamma\ket{x_0A}\bra{x_0A}}.
\end{equation}
With the bare Green's function $G(\omega)=(i\omega-X)^{-1}$, the matrix elements of the modified Green's function are given by
\begin{equation}
	\begin{split}
	\braket{xB|\tilde G(\omega)|x_0A} &=\braket{xB|G(\omega)|x_0A}\left(\sum_{n=0}^{\infty}\left(\delta\gamma\braket{x_0A|G(\omega)|x_0A}\right)^n\right)\\
 &=\frac{\braket{xB|G(\omega)|x_0A}}{1-\delta\gamma\braket{x_0A|G(\omega)|x_0A}}.
	\end{split}
\end{equation}

It should be noted that the element $\braket{xB|\tilde G(\omega)|x_0A}$ is only changed by a factor $N(\omega)=(1-\delta\gamma\braket{x_0A|G(\omega)|x_0A})^{-1}$ that is independent of the spatial coordinate $x$. As a result, $\braket{xB|\tilde G(\omega)|x_0A}$ and $\braket{xB|G(\omega)|x_0A}$ share the same spatial profiles as long as $N(\omega)$ is finite. The imbalance-induced factor $N(\omega)$ changes the coefficients $f_{R/L}(\omega)$ in Eq.  \eqref{seq:bulk_B_integral}, but leaves the scaling parts $|\beta_{R/L}(\omega)|^{2(x-x_0)}$ unaffected.

We consider $0<t_1<t_2$ below, where the dissipative PBC spectrum of $X$ (and $\tilde X$) is gapless. As shown in the last section, we have obtained 
\begin{equation}
	|\braket{xB|G(\omega_0+\delta\omega)|x_0A}|^2\sim (\delta\omega)^2e^{-2K(\delta\omega)^2|x-x_0|}
\end{equation}
when $x<x_0$. We can also get  \cite{xue2021simple} 
\begin{equation}
\begin{split}
	\braket{x_0A|G(\omega)|x_0A}&=\oint_{\text{GBZ}} \frac{\mathrm{d} \beta}{2 \pi i \beta} \left(\frac{1}{i\omega+ 0^{+}-X(\beta)}\right)_{AA}\\
 &=\frac{i\omega-\frac{t_2}{2}(\beta_R-\beta_R^{-1})+\gamma_1}{t_2(t_1+\frac{\gamma}{2})(\beta_R-\beta_L)}+\frac{1}{2t_1-\gamma}.
 \end{split}\label{seq:A_site_integral}
\end{equation}
To obtain this result, we use the fact that the integrand $
	\frac{1}{\beta} \left(\frac{1}{i\omega+ 0^{+}-X(\beta)}\right)_{AA}=\frac{\gamma_1+i\omega-\frac{t_2}{2}(\beta-\beta^{-1})}{\beta\det[i\omega+0^+-X(\beta)]}$ has two poles $\beta=0$ and $\beta=\beta_R(\omega)$ inside the GBZ. 

Because the integral in Eq.  \eqref{seq:bulk_B_integral} is dominated by the neighbor of $\omega_0$, we can expand $\braket{x_0A|G(\omega)|x_0A}$ as a series of $\delta\omega=\omega-\omega_0$ to obtain $\braket{x_0A|G(\omega)|x_0A}=\braket{x_0A|G(\omega_0)|x_0A}+K_1\delta\omega+\cdots$. For simplicity, we consider $\omega_0=\sqrt{t_2^2-t_1^2}$ here, while another gapless point $\omega_0=-\sqrt{t_2^2-t_1^2}$ provides a similar result. With the aforementioned expressions of $\beta_{L/R}(\omega_0)$, a direct calculation shows that 
\begin{equation}
	\begin{split}
\braket{x_0A|G(\omega_0)|x_0A}&=\frac{t_1+\gamma_1}{t_1(2t_1+\gamma_1)},\\
K_1=\left.\frac{\mathrm{d}\braket{x_0A|G(\omega)|x_0A}}{\mathrm{d}\omega}\right|_{\omega=\omega_0}&=\frac{\gamma_1  \sqrt{t_{2}^{2}-t_{1}^{2}}}{t_{1}^{3}\left( \gamma_1+2i \sqrt{t_{2}^{2}-t_{1}^{2}}\right)}.
 \end{split}
\end{equation}

\begin{figure}
	\centering
	\includegraphics[width=8.5cm]{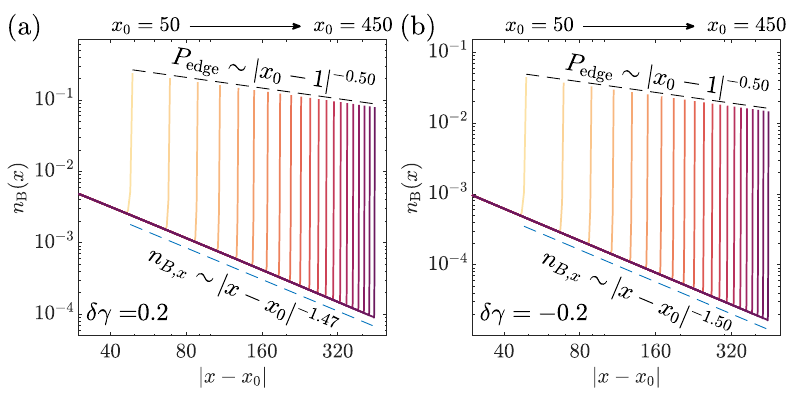}
	\caption{The steady-state edge burst in the presence of imbalanced gain and loss. $n_B(x)$ is steady-state particle density at $B$ sites. The pumping site $x_0$ ranges from $x_0=50$ to $x_0=450$. $\gamma_g=1.0$ in (a) and $\gamma_g=0.6$ in (b). Other parameters: $t_1=0.5,\ t_2=1,\ \gamma_1=\gamma_l=0.8,\ L=500$. Under these parameters, the critical imbalance is $\delta\gamma_c=9/13$.}
	\label{sfig:imbalance}
\end{figure}

If $1-\delta\gamma\braket{x_0A|G(\omega_0)|x_0A}\ne0$ for a small imbalance $\delta\gamma$, the zeroth-order approximation $N(\omega_0)$ attached to $f_{R/L}(\omega_0+\delta\omega)$ does not affect the asymptotic profiles of the integral in Eq.  \eqref{seq:bulk_B_integral}. Therefore, the steady-state bulk density $n_B(x)$ still exhibits an algebraical decay, and the steady-state edge burst still presents [Fig.  \ref{sfig:imbalance}].

Since $\braket{x_0A|G(\omega_0)|x_0A}$ is a real number, there exists a critical $\delta\gamma_c$ that makes $1-{\delta\gamma_c}\braket{x_0A|G(\omega_0)|x_0A}=0$. In this case, the divergence of $N(\omega_0)$ indicates that the eigenvalue of the impurity state under OBC touches the imaginary axis exactly at $\lambda_0=\pm i\sqrt{t_2^2-t_1^2}$. It also means that the system under OBC becomes unstable when $\delta\gamma>\delta\gamma_c$. Consequently, there does not exist a well-defined steady state where the number of particles is expected to be finite. The critical imbalance is given by
\begin{equation}
	\delta\gamma_c=\frac{1}{\braket{x_0A|G(\omega_0)|x_0A}}=\frac{t_1(2t_1+\gamma)}{t_1+\gamma}.
\end{equation}

In conclusion, as long as $\delta\gamma<\delta\gamma_c$, the steady-state edge burst remains robust in quadratic bosonic open quantum systems with local imbalanced gain and loss.

To end this part, we briefly remark on the steady-state edge burst in quadratic fermionic open quantum systems. In such systems, each site has a fermionic mode, and the anticommuting relation between them is given by $\{c_i^\dagger,c_j^\dagger\}=\{c_i,c_j\}=0, \{c_i^\dagger,c_j\}=\delta_{ij}$. In the calculation of the fermionic two-point correlators, the fermionic damping matrix is given by
 \begin{equation}
 	X_F=iH_0^T-\left(M^{(g)}+(M^{(l)})^T\right),
 \end{equation}
where $M^{(g)}=(D^{(g)})^\dagger D^{(g)}$ and  $M^{(l)}=(D^{(l)})^\dagger D^{(l)}$. The matrices $D^{(g)}$ and $D^{(l)}$ are defined by the single-particle fermionic gain and loss operators: $L^{(g)}_\mu=\sum_{i}D^{(g)}_{\mu,i}c^\dagger_i$ and $L^{(l)}_\mu=\sum_{i}D^{(l)}_{\mu,i}c_i$. The important feature is that the fermionic damping matrix $X_F$ is always stable. 

Specifically, we consider a fermionic open quantum system with the same setups as in Fig. \ref{fig:mapping}(c) of the main text. The fermionic damping matrix  becomes
\begin{equation}
	X_F=X-(\gamma_g+\gamma_l)|x_0A\rangle\langle x_0A|,\label{fermionic_damping matrix}
\end{equation}
where $X$ is equal to the balanced bosonic damping matrix in Eq.  \eqref{seq:balanced_damping_matrix}. Interestingly, the fermionic damping matrix $X_F$ is similar to the imbalanced bosonic case $\tilde{X}$ in Eq.  \eqref{seq:imbalanced_damping_matrix} with $\delta\gamma=-(\gamma_g+\gamma_l)<0<\gamma_c$.  Therefore, it is impossible to have unstable impurity states in fermionic systems. Following the above discussion, it is easy to find that the steady-state edge burst always exists in fermionic systems, as long as $X_F$ has NHSE and its dissipative PBC spectrum is gapless [which are indeed fulfilled by the $X_F$ in Eq.  \eqref{fermionic_damping matrix}]. 

\section{ Positive-$P$ method}\label{ap:positive-P}
In this section, we will provide a self-content tutorial for the positive-$P$ method. It is a numerical approach to simulate bosonic many-body open quantum systems. The main idea of this method is to map a bosonic density matrix to a probability distribution function in phase space, where the Lindblad master equation is transferred into a Fokker-Planck equation. This Fokker-Planck equation can be efficiently simulated by a corresponding stochastic process in the phase space.  For concreteness, we show how this method works in the bosonic open quantum system that consists of a quadratic Hamiltonian $\mathcal{H}_0=\sum_{ij}b_i^\dagger (H_0)_{ij}b_j$, linear Lindblad operators $L_\mu^{(g)}=\sum_iD^{(g)}_{\mu,i}b_i^\dagger$ and $L_\mu^{(l)}=\sum_jD^{(l)}_{\mu,j}b_j$, and two-body loss operators $L_{2,i}=\sqrt{\Gamma_i}b_ib_i$. (The indices $i,j$ mark the site positions on a one-dimensional chain.) The model considered in the main text belongs to this kind.

Following the standard procedure of the positive-$P$ method  \cite{deuar2021fully}, we can represent the density matrix of $N$ bosonic modes as a probability distribution function in a $2N$-dimensional complex phase space
\begin{equation}
	{\rho}(t)=\int \mathrm{d}^{2 N} \boldsymbol{\alpha} \mathrm{d}^{2 N} {\boldsymbol{\beta}} P\left(\boldsymbol{\alpha}, {\boldsymbol{\beta}},t\right) {\Lambda}\left(\boldsymbol{\alpha}, {\boldsymbol{\beta}}\right),\label{seq:P_rep}
\end{equation}
where $\boldsymbol{\alpha}=\{\alpha_1,\cdots,\alpha_N\}$ and $\boldsymbol{\beta}=\{\beta_1,\cdots,\beta_N\}$ are $2N$ independent complex phase-space variables and $\mathrm{d}^{2 N} \boldsymbol{\alpha} \mathrm{d}^{2 N}\boldsymbol{\beta}$ is a $4N$-dimensional real integral. In the above, ${\Lambda}(\boldsymbol{\alpha}, {\boldsymbol{\beta}})=\bigotimes_{i} {\Lambda}_{i}(\alpha_{i}, \beta_{i})$ is the double-phase-space basis of the positive-$P$ representation, where ${\Lambda}_{i}(\alpha_{i}, \beta_{i})=\frac{\ket{\alpha_{i}}\bra{\beta^*_{i}}}{\braket{\beta^*_{i}|\alpha_{i}}}=e^{-\alpha_i\beta_i}\ket{\alpha_{i}}\bra{\beta^*_{i}}$ is formed by the bosonic coherent states $\ket{\alpha_i}=e^{\alpha_{i}b_i^\dagger}\ket{\text{vac}}$ and $ \bra{\beta^*_i}=\bra{\text{vac}}e^{\beta_{i}b_i}$. We take the normalization $\operatorname{Tr}{\Lambda_{i}}=1$. $P(\boldsymbol{\alpha},\boldsymbol{\beta},t)$ is the function of probability distribution in the phase space, encoding the information of the density matrix $\rho(t)$. In the basis of $\Lambda({\boldsymbol{\alpha},\boldsymbol{\beta}})$, $P(\boldsymbol{\alpha},\boldsymbol{\beta},t)$ can be made positive real everywhere for any density matrix  \cite{Drummond_1980}. 

With the representation Eq.  \eqref{seq:P_rep}, the operations acting on the density matrix in the Lindblad master equation are converted to the corresponding multipliers and derivatives acting on the coherent basis ${\Lambda}(\boldsymbol{\alpha}, {\boldsymbol{\beta}})$. This can be done by utilizing the following identities:
\begin{equation}
	\begin{split}
		b_{i}  {\Lambda}(\boldsymbol{\alpha},\boldsymbol{\beta})=\alpha_{i}{\Lambda}(\boldsymbol{\alpha},\boldsymbol{\beta}),\quad &b_{i}^{\dagger}\Lambda(\boldsymbol{\alpha},\boldsymbol{\beta})=\left[\beta_{i}+\frac{\partial}{\partial \alpha_{i}}\right]  {\Lambda}(\boldsymbol{\alpha},\boldsymbol{\beta}), \\
        {\Lambda}(\boldsymbol{\alpha},\boldsymbol{\beta}) b_{i}^{\dagger}=\beta_{i} {\Lambda}(\boldsymbol{\alpha},\boldsymbol{\beta}),\quad&{\Lambda}(\boldsymbol{\alpha},\boldsymbol{\beta}) b_{i}=\left[\alpha_{i}+\frac{\partial}{\partial \beta_{i}}\right]  {\Lambda}(\boldsymbol{\alpha},\boldsymbol{\beta}).
	\end{split}
\end{equation}
These identities stem from the definition of bosonic coherent states. Based on them, we can rewrite the Lindblad master equation in the language of positive-$P$ representation ${\rho(t)}=\int \mathrm{d}^{4 N} \vec{v} P(\vec{v},t)  {\Lambda}(\vec{v})$ where $\vec v=\{\boldsymbol{\alpha},\boldsymbol{\beta}\}$.

As a result, the time evolution of the density matrix under the Lindblad master equation is equivalent to 
\begin{equation}
	\begin{aligned}
		\int \mathrm{d}^{4N} \vec{v}  &\frac{\partial P}{\partial t}{\Lambda} =\int \mathrm{d}^{4N} \vec{v} P\left\{\sum_{v} A_{v}(\vec{v}) \frac{\partial}{\partial v}+\sum_{v v^{\prime}} \frac{D_{v v^{\prime}}(\vec{v})}{2} \frac{\partial^{2}}{\partial v \partial v^{\prime}}\right\}  {\Lambda}\\	
		&=\int \mathrm{d}^{4N} \vec{v}  {\Lambda}\left\{-\frac{\partial}{\partial v} \sum_{v} A_{v}(\vec{v})+\sum_{n, \prime} \frac{\partial^{2}}{\partial v^{\prime} \partial v} \frac{D_{v v^{\prime}}(\vec{v})}{2}\right\} P.
	\end{aligned}	\label{seq:fokker-planck1}
\end{equation}
In the second line, we perform integration by parts and assume that boundary terms vanish at $|\alpha_i|,\ |\beta_i|\to+\infty$. The drift coefficients $A_v(\vec v)$ and the diffusion matrix $D_{vv'}(\vec v)$ are model-dependent. Their specific forms for the model in which we are interested in this section are
\begin{equation}
	\begin{split}
	A_j(\vec{v})&=\begin{pmatrix}
		\sum_iX^*_{ji}\alpha_i-2{\Gamma_j}\alpha_j^2\beta_j\\
		\sum_iX_{ji}\beta_i-2{\Gamma_j}\alpha_j\beta_j^{2}
	\end{pmatrix};\\
	D_{ij}(\vec{v})&=\begin{pmatrix}
		-2\Gamma_i \alpha_i^2\delta_{ij} & 2M^{(g)}_{ji}\\
		2M^{(g)}_{ij}&-2\Gamma_j\beta_j^{2}\delta_{ij}
	\end{pmatrix},
\end{split}
\end{equation}
where $X=iH_0^T+\sum_\mu (D^{(g)}_\mu)^\dagger D^{(g)}_\mu-\sum_\mu (D^{(l)}_\mu)^T (D^{(l)}_\mu)^* $ and $M^{(g)}=\sum_\mu (D^{(g)}_\mu)^\dagger D^{(g)}_\mu$. 

A direct solution to Eq.  \eqref{seq:fokker-planck1} yields a Fokker-Planck equation for $P(\boldsymbol{\alpha},\boldsymbol{\beta},t)$:
\begin{equation}
	\frac{\partial P}{\partial t}=\left\{-\frac{\partial}{\partial v} \sum_{v} A_{v}(\vec{v})+\sum_{n, \prime} \frac{\partial^{2}}{\partial v^{\prime} \partial v} \frac{D_{v v^{\prime}}(\vec{v})}{2}\right\}P.
\end{equation}
It is known that the Fokker-Planck equation is equivalent to a set of stochastic differential equations in phase space $\vec v=\{\boldsymbol{\alpha},\boldsymbol{\beta}\}$  \cite{carmichael1999statistical,gardiner2004quantum}. These stochastic differential equations are
\begin{equation}
	\frac{\partial \vec{v}}{\partial t}=A(\vec{v})+B(\vec{v})\vec{\xi}(t),
\end{equation}
where $\vec{\xi}(t)$ are $2N$ uncorrelated real Gaussian white noises with $\langle\xi_{v}(t)\rangle=0$ and $\langle\xi_{v}(t) \xi_{v^{\prime}}\left(t^{\prime}\right)\rangle=\delta\left(t-t^{\prime}\right) \delta_{v v^{\prime}} $. The matrix $B(\vec v)$ is defined such that $D=BB^T$. Numerically, these stochastic differential equations can be simulated by
\begin{eqnarray}
	\mathrm{d} \vec{v}=A(\vec{v})\mathrm{d}t+B(\vec{v})\mathrm{d}		\overrightarrow{W}.\label{seq:general_stochastic}
\end{eqnarray}
$\mathrm{d}\overrightarrow{W}=\vec{\xi}(t)\mathrm{d}t$ is the standard Wiener increments at each time step of length $\mathrm{d}t$, which are implemented by Gaussian random variables with variance $\mathrm{d}t$. Now the physical observables can be calculated in a stochastic manner. For example, the two-point correlator  $\Delta_{ij}(t)=\langle {b}_{i}^{\dagger}{b}_{j}\rangle=\operatorname{Tr}[\rho (t){b}_{i}^{\dagger}  {b}_{j} ]$ and the four-point correlator $C_{ijkl}(t)=\langle {b}_{i}^{\dagger}  {b}_{j}^{\dagger} {b}_{k}  {b}_{l}\rangle=\operatorname{Tr}[ \rho(t){b}_{i}^{\dagger}  {b}_{j}^{\dagger}   {b}_{k} {b}_{l} ]$ can be expressed as 
\begin{equation}
\begin{split}
	\Delta_{ij}(t)
	&	= \int \mathrm{d}^{4N} \vec{v} P(\vec{v},t) \beta_{i} \alpha_{j} 
		= \lim _{\mathcal{S} \to+\infty}\langle\langle \beta_{i}(t)\alpha_{j}(t)\rangle\rangle,\\
  C_{ijkl}(t) &=\int \mathrm{d}^{4N} \vec{v} P(\vec{v},t)\beta_{i} \beta_{j} \alpha_{k}\alpha_{l} =\lim _{\mathcal{S} \to \infty}\langle\langle \beta_{i}(t)\beta_{j}(t)\alpha_{k}(t)\alpha_{l}(t)\rangle\rangle .
  \end{split}
	\label{seq:stochastic_two_correlator}
\end{equation}
In the above, $\braket{\braket{\cdots}}$ represents the average over stochastic trajectories generated by Eq.  \eqref{seq:general_stochastic} and $\mathcal{S}$ denotes the total number of samples. While only an infinite $\mathcal{S}$ can make Eq. 
  \eqref{seq:stochastic_two_correlator} exact, a finite $\mathcal{S}$ is often sufficient to obtain reliable results. Furthermore, steady-state correlators can be achieved by taking the limit $t\rightarrow \infty$ in Eq.  \eqref{seq:stochastic_two_correlator}.

In summary, the application of the positive-$P$ method includes these key steps: (i) finding the drift coefficients $A_v(\vec v)$ and the diffusion matrix $D_{vv'}(\vec v)$ for the model; (ii) generating stochastic samples according to Eq. 
  \eqref{seq:general_stochastic}; (iii) calculating the correlators by averaging the samples. In our stochastic simulation of Figs. \ref{fig:P_rep_scaling}(a)--(d), we set the initial conditions as $\alpha_i(0)=\beta_i(0)=0$. The time step is $\mathrm{d}t=2\times 10^{-3}$ and we extract steady states at $t=200$. Due to a redundancy in the definition of the $B$ matrix, we take the following form in our simulation:

\begin{equation}
	\begin{split}
	B_{ij}(\vec v)=&\sum_x\delta_{i,xB}\delta_{j,xB}\begin{pmatrix}
		\sqrt{-2\gamma_2\alpha_i^2} &0\\
		0& \sqrt{-2\gamma_2\beta_i^2}	
	\end{pmatrix}\\
	&+ 	\sqrt{\frac{\gamma_g}{2}}\delta_{i,x_0A}\delta_{j,x_0A}\begin{pmatrix}
		1+i&1-i\\ 1-i &1+i
	\end{pmatrix},
\end{split}
\end{equation}

\section{Mean-field approximation in dissipative many-body systems}\label{ap:meanfield}
In this section, we provide a detailed discussion of the mean-field approximation in systems with two-body loss.  To do this, we introduce the two-point correlator $\Delta_{ij}(t)=\langle {b}_{i}^{\dagger}{b}_{j}\rangle=\operatorname{Tr}[\rho (t){b}_{i}^{\dagger}  {b}_{j} ]$ and the four-point correlator $C_{ijkl}(t)=\langle {b}_{i}^{\dagger}  {b}_{j}^{\dagger} {b}_{k}  {b}_{l}\rangle=\operatorname{Tr}[ \rho(t){b}_{i}^{\dagger}  {b}_{j}^{\dagger}   {b}_{k} {b}_{l} ]$ with $i,j,k,l$ denoting $\{(x,s)\}$. We focus on the balanced case $\gamma_g=\gamma_l$ in this section.

By utilizing the Lindblad master equation under the conditions $\gamma_1=0$ and $\gamma_2\ne0$, we can derive the hierarchy of time evolution of these correlators:
\begin{equation}
	\begin{split}
		\frac{\mathrm{d}\Delta_{xA,yA}(t)}{\mathrm{d}t}=&\left(X_0\Delta(t)+\Delta(t) X_0^\dagger+2M^{(g)}\right)_{xA,yA},\\
		\frac{\mathrm{d}\Delta_{xA,yB}(t)}{\mathrm{d}t}=&\left(X_0\Delta(t)+\Delta(t) X_0^\dagger+2M^{(g)}\right)_{xA,yB}-2\gamma_2C_{xA, yB, yB,yB}(t),\\
		\frac{\mathrm{d}\Delta_{xB,yA}(t)}{\mathrm{d}t}=&\left(X_0\Delta(t)+\Delta(t) X_0^\dagger+2M^{(g)}\right)_{xB,yA}-2\gamma_2C_{xB, xB, xB, yA}(t),\\
		\frac{\mathrm{d}\Delta_{xB,yB}(t)}{\mathrm{d}t}=&\left(X_0\Delta(t)+\Delta(t) X_0^\dagger+2M^{(g)}\right)_{xB,yB}\\
  &\quad-2\gamma_2\left(C_{xB, xB, xB, yB}(t)+C_{xB, yB, yB, yB}(t)\right).\\
	\end{split}\label{seq:delta_two_body_evolution}
\end{equation}
In the above, $X_0\equiv X|_{\gamma_1=0}$ where $X$ is given by Eq.  \eqref{seq:balanced_damping_matrix} under the condition $\gamma_g=\gamma_l$. 

\begin{figure}
	\centering
 \includegraphics[width=8.5cm]{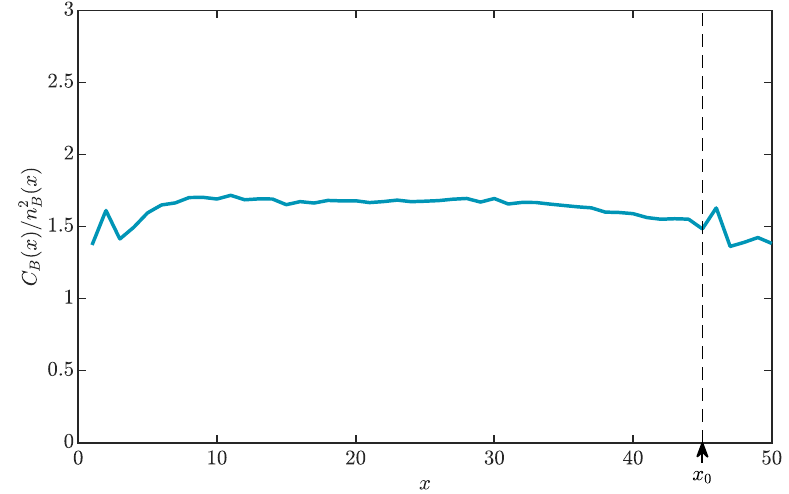}
	\caption{The ratio $C_B(x)/n_B^2(x)$ of the steady state in two-body loss systems. We take the steady values $C_B(x)$ and $n_B(x)$ from Figs. \ref{fig:P_rep_scaling}(a) and 3(b) of the main text.}
	\label{sfig:four_to_two_ratio}
\end{figure}
The numerical simulation based on the positive-$P$ method yields an approximated relation 
\begin{equation}
	\frac{C_{xB,xB,xB,xB}(t\to+\infty)}{\Delta_{xB,xB}(t\to+\infty)^2}\approx O(1)
\end{equation} of the steady state. The result is shown in Fig.  \ref{sfig:four_to_two_ratio}, where we define $C_B(x)=C_{xB,xB,xB,xB}(t\to+\infty)$ and $n_B(x)=\Delta_{xB,xB}(t\to+\infty)$ as the steady values. This approximated relation motivates us to consider the following substitutions: 
\begin{equation}
	\begin{split}
	C_{xB,xB,xB,yB}(t)\approx\Delta_{xB,xB}(t)\Delta_{xB,yB}(t);\\
	C_{xB,yB,yB,yB}(t)\approx\Delta_{xB,yB}(t)\Delta_{yB,yB}(t);\\
	C_{xB,xB,xB,yA}(t)\approx\Delta_{xB,xB}(t)\Delta_{xB,yA}(t);\\
	C_{xA,yB,yB,yB}(t)\approx\Delta_{xA,yB}(t)\Delta_{yB,yB}(t).
	\end{split}
\end{equation}

Since the density at the $B$ sites is defined as $n_{B}(x,t)=\Delta_{xB,xB}(t)$, putting these relations back into Eq.  \eqref{seq:delta_two_body_evolution} gives rise to
\begin{equation}
	\frac{\mathrm{d}}{\mathrm{d}t}\Delta_{\text{MF}}(t)=X_{\text{MF}}(t)\Delta_{\text{MF}}(t)+\Delta_{\text{MF}}(t)X_{\text{MF}}(t)^\dagger+2M^{(g)},\label{seq:mean_field_evolution}
\end{equation}
where the mean-field damping matrix $X_{\text{MF}}$ in real space is given by replacing $X_{xB,xB}=-\gamma_1$ in Eq.  \eqref{seq:balanced_damping_matrix} with
\begin{equation}
	[X_{\text{MF}}(t)]_{xB,xB}=-2\gamma_2n_{B,\text{MF}}(x,t)=-2\gamma_2[\Delta_{\text{MF}}(t)]_{xB,xB}.\label{seq:mean_field_damping_matrix}
\end{equation}  
This mean-field substitution introduces a nonlinear evolution equation to the mean-field two-point correlator $\Delta_{\text{MF}}(t)$. The steady-state results at the mean-field level, shown in Fig. \ref{fig:P_rep_scaling}(e,f) of the main text, are obtained by simulating this nonlinear equation from the null initial state $\Delta_{\text{MF}}(0)=0$  for a sufficiently long time.

Intuitively, the mean-field damping matrix $X_{\text{MF}}$ is equivalent to employing the density-dependent single-body loss operator
\begin{eqnarray}
L_{\text{MF},xB}=\sqrt{2\gamma_2n_{B,\text{MF}}(x,t)}b_{xB}.\label{seq:mean_field}
\end{eqnarray}
It qualitatively imitates the two-body loss operator $L_{2,xB}=\sqrt{\gamma_2}b_{xB}b_{xB}$ at the mean-field level. The prefactor $2$ arises from the operator exchange during the derivation of Eq.  \eqref{seq:delta_two_body_evolution}, but it does not influence the discussion of the steady-state edge burst and bulk-edge scaling relation. Therefore, we have justified the validity of the mean-field approximation employed in the main text.

The mean-field approximation provides an intuitive picture of the dissipative many-body system. In the case of $0<t_1<t_2$ where the condition $t_1+t_2\cos k_0=0$ can be satisfied, the mean-field damping matrix $X_{\text{MF}}(t)$ on a circular chain always possesses some gapless eigenmodes, which are related to the dark states $\ket{\phi_n}=\frac{1}{\sqrt{n!}}(b_{k_0}^\dagger)^n\ket{0}$ mentioned in the main text. These dark states indicate the absence of a dissipative gap under PBC. Consequently, particles  incoherently pumped into the system from site $A$ at $x_0$ can travel long distances along the $A$ chain before escaping to the environment from sites $B$ or being scattered by the boundaries. This intuitive picture supports the notion of long-range algebraic decay of steady-state correlators on $B$ sites.

The mean-field theory shows that the loss strength at $x$ is determined by the local particle density $2\gamma_2 n_{B,\text{MF}}(x)$.  It is important to note that $2\gamma_2n_{B,\text{MF}}(x)$ in the bulk region $(1\ll x\ll x_0)$ is very small compared to the values of $t_1$ and $t_2$ [Fig. \ref{fig:P_rep_scaling}(f) in the main text].  From a local perspective, particles traveling along the dissipationless $A$ chain do not suffer the loss effect on the $B$ sites until they jump to the $B$ chain at $x$. In the vicinity of $x$, we can approximately treat particle motion as occurring in a uniform system with a local mean-field damping matrix $\tilde X_{\text{MF}}(x)\equiv \left.X\right|_{\gamma_1\to 2\gamma_2n_{B,\text{MF}}(x)}$. $X$ is given by Eq.  \eqref{seq:balanced_damping_matrix}. Interestingly, $\tilde X_{\text{MF}}(x)$ also exhibits gapless states in a periodic system and has NHSE under OBC. To capture the effect of the small density-dependent loss, we need to reconsider the expansion presented in Eq.  \eqref{seq:delta_omega_expansion} for both a small $\delta\omega$ and a small $\tilde\gamma\equiv\gamma_1\to2\gamma_2n_{B,\text{MF}}(x)$:
\begin{equation}
		\begin{split}
		f_L(\omega_0+\delta\omega)&\sim(\tilde\gamma)^0(\delta\omega)^1,\\
		\ln|\beta_L(\omega_0+\delta\omega)|&\sim(\tilde\gamma)^1(\delta\omega)^2.
	\end{split}
\end{equation}
Here, we use the fact that $K$ defined below Eq.  \eqref{seq:delta_omega_expansion} is proportional to $\tilde\gamma\equiv\gamma_1$ when $\tilde\gamma$ is very small. These expansions lead to
\begin{equation}
	\begin{split}
		n_{B,\text{MF}}(x)\sim\int\mathrm{d}\omega(\delta\omega)^2e^{-2K'\tilde\gamma(\delta\omega)^2|x-x_0|}\sim(\tilde\gamma)^{-\frac{3}{2}}|x-x_0|^{-\frac{3}{2}},
	\end{split}
\end{equation}
where $K'$ is an irrelevant constant that is independent of $\tilde \gamma$. With $\tilde\gamma\sim2\gamma_2n_{B,\text{MF}}(x)$, we can obtain 
\begin{equation}
	n_{B,\text{MF}}(x)\sim(n_{B,\text{MF}}(x))^{-\frac{3}{2}}|x-x_0|^{-\frac{3}{2}},
\end{equation}
which leads to
\begin{equation}
	n_{B,\text{MF}}(x)\sim|x-x_0|^{-\frac{3}{5}}=|x-x_0|^{-0.6}.
\end{equation}

Remarkably, the mean-field approximation in the two-body loss system leads to an algebraic decay of the steady-state bulk density $ n_{B,\text{MF}}(x)$. Additionally, this scaling behavior gives rise to $ n_{B,\text{MF}}^2(x)\sim|x-x_0|^{-1.2}$.

At the mean-field level, the steady-state condition $\frac{\mathrm{d}N_{\text{tot}}}{\mathrm{d}t}=0$ provides a constraint as follows:
\begin{equation}
	\sum_{x=1}^Ln_{B,\text{MF}}^2(x)=\frac{\gamma_g}{2\gamma_2}.
\end{equation}
As explained in the main text, in a finite system with boundaries, many-body NHSE induced by two-body loss causes the remaining particles outside of this system to accumulate predominantly at the left boundary. This accumulation results in
\begin{equation}
n_{B,\text{MF},\text{edge}}^2=n^2_{B,\text{MF}}(x=1)\sim\int_{-\infty}^1\mathrm{d}x|x-x_0|^{-1.2}\sim|x_0-1|^{-0.2},
\end{equation}
which further provides $n_{B,\text{MF},\text{edge}}\sim|x_0-1|^{-0.1}$. 

In conclusion, by applying the mean-field approximation, we have successfully identified the scaling factors associated with the bulk and edge values of $n_{B,\text{MF}}(x)$ and $n_{B,\text{MF}}^2(x)$. Comparing these theoretical results with the numerical findings presented in Figs. \ref{fig:P_rep_scaling}(c-f) of the main text, we observe a reasonable agreement between the mean-field estimations and the numerical results. Additionally, the scaling factors derived from the mean-field analysis align well with the results obtained by the positive-$P$ method. These results demonstrate the bulk-edge scaling relations of the steady-state edge burst in dissipative many-body systems. In the dissipative many-body system with two-body loss, we obtain two bulk-edge scaling relations: $\alpha_e=\alpha_b-1$ for the steady-state four-point correlators $C_B(x)$ and $\alpha_e=\alpha_b-0.5$ for the steady-state density distribution $n_B(x)$.

\bibliography{dirac}

\end{document}